\documentclass{article}%
\usepackage{amsmath}
\usepackage{amsfonts}
\usepackage{amssymb}
\usepackage{graphicx}%
\providecommand{\U}[1]{\protect\rule{.1in}{.1in}}
\providecommand{\U}[1]{\protect\rule{.1in}{.1in}}
\newtheorem{theorem}{Theorem}

\newtheorem{proposition}[theorem]{Proposition}

\begin{document}

\author{ \textbf{\ Demosthenes Ellinas} \thanks{\texttt{ellinas@science.tuc.gr}}\\Department of Sciences, Division of Mathematics, \\Technical University of Crete, GR 731 00 Chania Crete Greece}
\title{\textbf{Constructing Operator Valued Probability Measures in Phase Space}}
\maketitle

\begin{abstract}
Probability measures (quasi probability mass), given in the form of integrals
of Wigner function over areas of the underlying phase space, give rise to
operator valued probability measures (OVM). General construction methods of
OVMs, are investigated in terms of geometric positive trace increasing maps
(PTI), for general 1D domains, as well as 2D shapes e.g. circles, disks.
Spectral properties of OVMs and operational implementations of their
constructing PITs are discussed.



\end{abstract}




An OVM is a map $K:\mathcal{F\rightarrow L}(H),$ from a $\sigma-$algebra
$\mathcal{F}$ of subsets of an nonempty set $\Omega,$ to the bounded operators
$\mathcal{L}(H)$ on a Hilbert space $H,$ such that for $\Psi\in H,$ and
$X\in\mathcal{F}$, \ the function $\mu_{\Psi}(X)\equiv\langle\Psi
|K(X)\Psi\rangle,$ is a normalized (i.e. $\mu_{\Psi}(\Omega)=1$), generalized
(i.e. negative valued) measure (i.e. $\sigma-$additive set function). \ If
$F(X)\geq0,$ i.e. $\mu_{\Psi}(X)>0,$ $\forall X\in\mathcal{F}$ \ we have a
positive OVM, further if $K(X)^{\dagger}=K(X),$ and $K(X)^{2}=K(X),$ we have a
projective OVM, namely an observable (see e.g. \cite{bgl}). For the case of
Wigner function\cite{schbook}\cite{royer} $W_{|\psi>}(\alpha)=\langle
\Psi|D(\alpha)\Pi D(\alpha)^{\dagger}|\Psi\rangle,$ we have $\Omega\equiv C,$
$H\equiv$span$(|n\rangle,n=0,1,...)$ the Fock space, with $N$ \ the number
operator, $\Pi=e^{i\pi N}$ the parity operator, and $D(\alpha)=e^{\alpha
a^{\dagger}-\alpha^{\ast}a}=e^{(iqP-ipQ)}$ the displacement operator.
Integrals of Wigner function\cite{bdw}\cite{bew} $\int_{X}W_{|\psi>}%
(\alpha)d^{2}\alpha=Tr(|\Psi\rangle\langle\Psi|K(X)),$ physically define the
\textit{quasi probability mass }over $X,$ in terms of OVM $K(X)\equiv\int
_{X}D(\alpha)\Pi D(\alpha)^{\dagger}d^{2}\alpha.$

\begin{proposition}
Let $X_{q}$ a region in $q$ - axis determined by the characteristic function
(cfun), $\chi(q),q\in R,$ with Fourier transform (FT) $\widetilde{\chi}(p),$
then the associated OVM is $K_{q}=\widetilde{\chi}(P)\Pi.$ The solution of
eigen problem $K_{q}|\Psi_{\pm}\rangle=\lambda_{\pm}|\Psi_{\pm}\rangle,$
determines the eigenvalues $\lambda_{\pm}(p)=\pm\widetilde{\chi}(p),$ and the
eigenvectors $|\Psi_{\pm}\rangle=|p\rangle\pm|-p\rangle.\ $If further
$\chi(q)$ is $L-$periodic with Fourier series $\chi(q)=\frac{a_{0}}{2}%
+\sum_{m=1}^{\infty}a_{m}\cos(\frac{m\pi q}{L})+b_{m}\sin(\frac{m\pi q}{L}),$
the OVM becomes
\begin{align}
K_{q}  &  =\Pi\left[  \frac{\pi a_{0}}{2}|p=0\rangle\langle p=0|+\pi\sum
_{m=1}^{\infty}r_{m}\left(  e^{i\phi_{m}}|p=\frac{m\pi}{L}\rangle\langle
p=\frac{m\pi}{L}|\right.  \right. \nonumber\\
&  \text{
\ \ \ \ \ \ \ \ \ \ \ \ \ \ \ \ \ \ \ \ \ \ \ \ \ \ \ \ \ \ \ \ \ \ \ \ \ \ \ \ \ }%
\left.  \left.  +e^{-i\phi_{m}}|p=-\frac{m\pi}{L}\rangle\langle p=-\frac{m\pi
}{L}|\right)  \right]  , \label{regoper}%
\end{align}
where $r_{m}\equiv|a_{m}+ib_{m}|$ ,and $\phi_{m}\equiv\arg(a_{m}+ib_{m}).$
\end{proposition}

\textit{Proof: \ }The \textit{region operator} (convention: alternative name
for OVM related to quasi-probability functions, see below), is constructed by
smearing along $q-$axis the point at the origin as follows%
\begin{equation}
K_{q}=\int_{X_{q}}e^{i\frac{qP}{2}}\Pi e^{-i\frac{qP}{2}}dq=\int_{R}%
\chi(q)e^{iqP}dq\Pi=\int_{R}\widetilde{\chi}(p)|p\rangle\langle-p|dp\equiv
\widetilde{\chi}(P)\Pi,
\end{equation}
\bigskip where $\widetilde{f}(p)=\int_{R}f(q)e^{iqp}dq,$ is the FT.
Straightforward evaluation of FT of the cfun of eq. (\ref{fourtrans}), below
leads to the result. The eigen problem is also easily solved utilizing the
action of parity operator on momentum state.$\blacksquare$

\textit{Remarks}:\textit{ }1) From FT of a $L-$periodic cfun of some region we
obtain the eigenvalues of its associateed OVM $\lambda_{\pm}(p)=\pm
\widetilde{\chi}(p),$ where
\begin{align}
\widetilde{\chi}(w)  &  =\allowbreak a_{0}\pi\delta\left(  w\right)
\nonumber\\
&  +\pi\sum_{m=1}^{\infty}\left[  a_{m}\left(  \delta\left(  w-\frac{m\pi}%
{L}\right)  +\delta\left(  w+\frac{m\pi}{L}\right)  \right)  -\allowbreak
ib_{m}\left(  \delta\left(  w-\frac{m\pi}{L}\right)  -\delta\left(
w+\frac{m\pi}{L}\right)  \right)  \right]  . \label{fourtrans}%
\end{align}
If cfun is asymmetric then its FT is complex valued, this implies the region
operator would not be a real one.

2) Let e.g. the pure density matrix $\rho=|\Psi\rangle\langle\Psi|,$
determined by an even parity state vector i.e. $P|\Psi\rangle=|\Psi\rangle,$
then the occupation probability of the region $X_{q}$ is
\begin{equation}
p_{\rho}(X_{q})=Tr(K_{q}\rho)=\int_{R}\widetilde{\chi}(p)|\Psi(p)|^{2}dp.
\end{equation}
If further the region $X_{q}$ is symmetric with respect to the origin, i.e.
its cfun is an even function ( i.e. $b_{m}=0$)$,$ then we obtain the
probability
\begin{equation}
p_{\rho}(X_{q})=a_{0}\pi|\Psi(0)|^{2}+2\pi\sum_{m=1}^{\infty}a_{m}|\Psi
(\frac{m\pi}{L})|^{2}.
\end{equation}

3) Let us choose the special region for which $a_{m}\geq0,$ $b_{m}=0,$ then
the associated region operator becomes
\begin{align}
K_{q}  &  =a_{0}\pi|p=0\rangle\langle p=0|+\pi\sum_{m=1}^{\infty}a_{m}\left(
|p=-\frac{m\pi}{L}\rangle\langle p=\frac{m\pi}{L}|\right. \nonumber\\
&  \left.  \text{
\ \ \ \ \ \ \ \ \ \ \ \ \ \ \ \ \ \ \ \ \ \ \ \ \ \ \ \ \ \ \ \ \ \ \ \ \ \ \ }%
+|p=\frac{m\pi}{L}\rangle\langle p=-\frac{m\pi}{L}|\right)  .
\end{align}
This is also expressed as$\ \ K_{q}=\varepsilon_{q}(|p=0\rangle\langle p=0|),$
with $\varepsilon_{q}$ been a PTI map is generated by the set of Kraus
generators\cite{kr} $\ \varepsilon_{q}\equiv\left\{  \sqrt{a_{0}\pi}%
\mathbf{1},\sqrt{a_{m}}e^{-i\frac{m\pi}{L}Q},\sqrt{a_{m}}e^{i\frac{m\pi}{L}%
Q}\right\}  _{m=1}^{\infty}.$ The latter expression of $K_{q},$ if utilized
together with state-observable duality (c.f. \cite{dual}), would provide, by
means of a unitarization of $\varepsilon_{q}$ map$,$ means for some physical
implementation of the general region operator $K_{q}$ (see similar
constructions of other region operators in \cite{ellinas}, and below).

4) The operator $K_{q}$ defined on a region of $q-$axis with cfun $\chi(q)$,
can be rotated by $\frac{\pi}{2}$ radians to become $K_{p},$ i.e. a similar
operator on $p$-axis which reads%
\begin{equation}
K_{p}\equiv e^{i\frac{\pi}{2}N}K_{q}e^{-i\frac{\pi}{2}N}=e^{i\frac{\pi}{2}%
N}\widetilde{\chi}(P)\Pi e^{-i\frac{\pi}{2}N}=\widetilde{\chi}(Q)\Pi.
\end{equation}
Having available\ operators $K_{q},$ $K_{p},$ been defined along orthogonal
axes, we further rotate them clockwise by an angle $\theta,$ to\ obtain the
respective operators \ along new rotated axes as follows,%
\begin{equation}
K_{q}^{\theta}=e^{i\theta N}K_{q}e^{-i\theta N}=e^{i\theta N}\widetilde{\chi
}(P)\Pi e^{-i\theta N}=\widetilde{\chi}(\cos\theta P-\sin\theta Q)\Pi,
\end{equation}
and
\begin{equation}
K_{p}^{\theta}=e^{i\theta N}K_{p}e^{-i\theta N}=e^{i\theta N}\widetilde{\chi
}(Q)\Pi e^{-i\theta N}=\widetilde{\chi}(\cos\theta Q+\sin\theta P)\Pi.
\end{equation}
These two operators are related by a $\frac{\pi}{2}$ radians rotation i.e.
$K_{p}^{\theta}=e^{i\frac{\pi}{2}N}K_{q}^{\theta}e^{-i\frac{\pi}{2}N}.$

5) Using properties of Fourier transform we obtain shift transforms of a
region operator e.g. $K_{q}e^{icP}=\widetilde{\chi}(P)\Pi$ as follows,%
\begin{equation}
e^{icP}K_{q}=\widetilde{\chi}(P+c\mathbf{1})\Pi,\text{ \ }K_{q}e^{icP}%
=\widetilde{\chi}(P-c\mathbf{1})\Pi,\text{ }e^{i\frac{c}{2}P}K_{q}%
e^{-i\frac{c}{2}P}=\widetilde{\chi}(P+c\mathbf{1})\Pi.
\end{equation}

6) Using the squeezing operator $S(\zeta)=e^{\zeta a^{\dagger2}-\zeta^{\ast
}a^{2}}=e^{\frac{1}{2}r(Q^{2}-P^{2})},$with parameter $\zeta=\frac{1}%
{2}re^{-i2\phi},$ we obtain by similarity transformation multiplicative
actions on position and momentum operators%
\begin{equation}
S(\zeta)^{\dagger}QS(\zeta)=Qe^{r},\text{\ }S(\zeta)^{\dagger}PS(\zeta
)=Pe^{-r}.
\end{equation}
This gives rise to a region operator with (squeezed) scaled support on
$q$-axis i.e.
\begin{equation}
S(\zeta)^{\dagger}K_{q}S(\zeta)=S(\zeta)^{\dagger}\widetilde{\chi}%
(P)S(\zeta)\Pi=\widetilde{\chi}(Pe^{-r})\Pi.
\end{equation}

7) OVM from $s$-Parametrized Quasiprobability Functions: Working out the power
series expression of $\ s-$parametrized quasi probabilities phase space
functions $F(\alpha;s),$ of \cite{mck,royer}
\begin{equation}
F(\alpha;s)=\frac{2}{\pi}\sum_{k=0}^{\infty}\frac{(1+s)^{k}}{(1-s)^{k+1}
}(-1)^{k}\langle k|D(\alpha)^{\dagger}\rho D(\alpha)|k\rangle,
\end{equation}
we obtain that
\begin{equation}
F(\alpha;s)=\frac{2}{\pi}Tr(\rho\widehat{F}(\alpha;s))\equiv\frac{2}{\pi
}Tr(\rho D(\alpha)\Pi(s)D(\alpha)^{\dagger}),
\end{equation}
where the $s-$parametrized parity operator $\Pi(s)=\frac{(1+s)^{N}
}{(1-s)^{N+1}}\Pi,$ is now displaced by the coherent state generating
operators. Special cases are the Glauber-Sudarshan $P$ function $F(\alpha
;s=1)=P(\alpha),$ the Wigner function $F(\alpha;s=0)=W(\alpha),$ and the $Q$
positive function $F(\alpha;s=-1)=Q(\alpha).$ The "point operator" at the
origin of phase plane is now the $s$-parametrized parity operator $\Pi(s).$
Applying appropriate geometric PTI maps $\Pi(s)$ as in the present case of
Wigner function, i.e. $s=0,$ we can derive $s$-parametrized OVMs.

\textit{Examples}: We proceed with examples that elucidate the general
construction: 1) Let us choose $X_{q}=\{1,2,...,n\},$ namely the case where
the region consists of the first $n$ positive integers. This set with cfun
$\chi(x)=1$ for $x=1,2,...,n,$ and $0$ otherwise, leads to region operator
\begin{equation}
K_{q}=\int_{R}\chi(x)e^{i2xP}dx\Pi=\sum_{m=1}^{n}e^{i2mP}\Pi=\frac
{e^{i(2n+1)P}-e^{iP}}{2i\sin P}\Pi.
\end{equation}
As the set is not symmetric with respect to the origin of axes, the resulting
region operator is a complex one in this case.

2) Circle OVM $K_{C}(a)$; construction steps: displace by $a$ along $q$-axis
the $0$ point OVM i.e. $\Pi\rightarrow e^{-iaP}\Pi e^{iaP},$ and rotate it
around by means of the continuous PTI map $\varepsilon_{2\pi},$ with Kraus
generators $\{e^{-i\phi N};0\leq\phi<2\pi\},$ i.e. $e^{-iaP}\Pi e^{iaP}%
\rightarrow\varepsilon_{2\pi}(e^{-iaP}\Pi e^{iaP}).$ Explicitly,%
\begin{align}
\Pi &  \rightarrow e^{-iaP}\Pi e^{iaP}\rightarrow\varepsilon_{2\pi}%
(e^{-iaP}\Pi e^{iaP})\equiv\int_{0}^{2\pi}d\phi e^{-i\phi N}e^{-iaP}\Pi
e^{iaP}e^{i\phi N}\nonumber\\
&  =\Pi\int_{0}^{2\pi}d\phi e^{-i\phi N}e^{2iaP}e^{i\phi N}=\Pi\sum
_{mn=0}^{\infty}\int_{0}^{2\pi}d\phi e^{-i\phi N}|m\rangle\langle
m|e^{2iaP}|n\rangle\langle n|e^{i\phi N}\\
&  =2\pi\Pi\sum_{n=0}^{\infty}e^{-a^{2}/2}L_{n}(a^{2})|n\rangle\langle
n|\equiv K_{C}(a),
\end{align}
where $\langle n|e^{2iaP}|n\rangle=e^{-a^{2}/2}L_{n}(a^{2}),$ $L_{n}$ been the
Laguerre polynomials (result obtained originally in \cite{bdw}).

3) Disc OVM $K_{D}(a)$; construction steps: start with OVM $K_{L}(a)$ of line
segment extended along $[-L,L],$ been constructed by means of PTI map in
\cite{ellinas}, now rotate it around by means of the continuous PTI map
$\varepsilon_{2\pi},$ with Kraus generators $\{e^{-i\phi N};0\leq\phi<2\pi\}.$
Explicitly,
\begin{align}
K_{L}(a)  &  =\frac{\sin(Pa)}{P}\Pi\rightarrow\varepsilon_{2\pi}
(K_{L}(a))\equiv\int_{0}^{2\pi}d\phi e^{-i\phi N}K_{L}(a)e^{i\phi
N}\nonumber\\
&  =\Pi\int_{0}^{2\pi}d\phi e^{-i\phi N}\frac{\sin(Pa)}{P}e^{i\phi N}
=\int_{-a/2}^{a/2}dx\left(  \Pi\int_{0}^{2\pi}d\phi e^{-i\phi N}
e^{2ixP}e^{i\phi N}\right) \\
&  =2\pi\Pi\sum_{n=0}^{\infty}\left(  \int_{-a/2}^{a/2}dxe^{-x^{2}/2}
L_{n}(x^{2})\right)  |n\rangle\langle n|=\int_{-a/2}^{a/2}dxK_{C}(x)\equiv
K_{D}(a).
\end{align}

\textit{Operational construction of OVMs}: Let $\Pi_{\alpha}^{1}=D_{\alpha}\Pi
D_{\alpha}^{\dagger}\otimes\mathbf{1,}\Pi_{\beta}^{2}=$ $\mathbf{1}\otimes
D_{\beta}\Pi D_{\beta}^{\dagger},$ be point operators with support on points
$\alpha,\beta\in C,$ belonging to two respective phase planes. Their sum
acting in $H\otimes H,$ is $\ \Pi_{\alpha}^{1}+\Pi_{\beta}^{2}=D_{\alpha
}\otimes D_{\beta}(\Pi^{1}+\Pi^{2})D_{\alpha}^{\dagger}\otimes D_{\beta
}^{\dagger}\mathbf{\ ,\ }$\ i.e. is the displaced sum of two operators with
support on the zero point of the respective phase planes. To construct the
latter we resort to the property: the unitary operator $V=\exp(-i\frac{\pi}%
{2}J),$ where $J=\frac{1}{2i}(a_{1}^{\dagger}a_{2}-a_{2}^{\dagger}a_{1}),$
serves as the permutation operators between parity operators of two different
Hilbert spaces i.e. $\Pi^{2}=V^{2}\Pi^{1}V^{\dagger2}$. Then $\Pi^{1}+\Pi
^{2}=\Pi^{1}+V^{2}\Pi^{1}V^{\dagger2}\equiv\varepsilon(\Pi^{1}),$ \ where we
have introduced the positive trace increasing map $\varepsilon,$ with two
Kraus operators $(\mathbf{1},V^{2})$. Let the density matrix $\rho$ acting in
$H\otimes H,$ and consider the expectation value $\langle\Pi^{1}+\Pi
^{2}\rangle\equiv Tr(\rho(\Pi^{1}+\Pi^{2}))$ $=Tr(\rho\varepsilon(\Pi
^{1}))=Tr(\varepsilon^{\ast}(\rho)\Pi^{1}),$ where the dual PTI map
$\varepsilon^{\ast}$ has been introduced (see \cite{dual}), as $\varepsilon
^{\ast}(\rho)=\rho+V^{\dagger2}\rho V^{2}.$ Obtaining the value $\langle
\Pi^{1}+\Pi^{2}\rangle,$ requires the operational construction of density
matrix $\varepsilon^{\ast}(\rho);$ which amounts to a unitary dilation of
$\varepsilon^{\ast}.\ $To accomplish this we introduce the auxiliary space
$H_{A}=$span$\{|0\rangle,|1\rangle\},$ and the unitary operator
\begin{equation}
W=\left(
\begin{array}
[c]{cc}%
\mathbf{1} & -V^{2}\\
V^{\dagger2} & \mathbf{1}%
\end{array}
\right)  ,
\end{equation}
acting on $H_{A}\otimes H\otimes H.$ Then we obtain $\varepsilon^{\ast}%
(\rho)=Tr_{A}W^{^{\dagger}}(|0\rangle\langle0|\otimes\rho)W;$ the partial
tracing map $Tr_{A}$ signifies operationally an unconditional measurment in
$H_{A}$\cite{fund}.

\vskip0.5cm \textit{Acknowledgment} : Work partially supported by EPEAEK:
Pythagoras II.


\begin{thebibliography}{99}
\bibitem {dual}By means of the cyclic property of trace, the expectation value
of an observable $X,$ viz. $\langle X,\varepsilon(\rho)\rangle=Tr(X\varepsilon
(\rho)),$ on a state $\rho,$ mapped by a CP map $\varepsilon,$ as
$\varepsilon(\rho)=\sum_{i}A_{i}\rho A_{i}^{\dagger},$ serves to define the
dual map $\ \varepsilon^{\ast}(X)=\sum_{i}A_{i}^{\dagger}XA_{i},$ that satisfy
the equation $\langle X,\varepsilon(\rho)\rangle=Tr(\rho\varepsilon^{\ast
}(X)).$
\end{thebibliography}
\end{document}